\documentclass[conference]{IEEEtran}
\IEEEoverridecommandlockouts
\usepackage{cite}
\usepackage{amsmath,amssymb,amsfonts}
\usepackage{algorithmic}
\usepackage{graphicx}
\usepackage{textcomp}
\usepackage{xcolor}
\usepackage{url}
\usepackage{fancyhdr,lipsum}
\def\BibTeX{{\rm B\kern-.05em{\sc i\kern-.025em b}\kern-.08em
    T\kern-.1667em\lower.7ex\hbox{E}\kern-.125emX}}

\makeatletter
 \let\old@ps@headings\ps@headings
 \let\old@ps@IEEEtitlepagestyle\ps@IEEEtitlepagestyle
 \def\confheader#1{%
 \def\ps@headings{%
 \old@ps@headings%
 \def\@evenhead{\strut\hfill#1\hfill\strut}%
 }%
 \def\ps@IEEEtitlepagestyle{%
 \old@ps@IEEEtitlepagestyle%
 \def\@oddhead{\strut\hfill#1\hfill\strut}%
 \def\@evenhead{\strut\hfill#1\hfill\strut}%
 }%
 \ps@headings%
 }
 \makeatother

\confheader{%
 2024 IEEE International Students' Conference on Electrical, Electronics and Computer Science
 }

 \usepackage[pscoord]{eso-pic}
\newcommand{\placetextbox}[3]{
 \setbox0=\hbox{#3}
 \AddToShipoutPictureFG*{ \put(\LenToUnit{#1\paperwidth},\LenToUnit{#2\paperheight}){\vtop{{\null}\makebox[0pt][c]{#3}}}
 }
 }
 \placetextbox{.23}{0.055}{\small{979-8-3503-4846-0/24/\$31.00 ©2024 IEEE}}
\fancypagestyle{plain}{

  \fancyfoot[C]{SCEECS 2024}

}

\begin{document}

\title{A survey of LSM-Tree based Indexes, Data Systems and KV-stores\\
}

\author{\IEEEauthorblockN{$Supriya Mishra^{1,2}$}
\IEEEauthorblockA{\textit{$^1$ Sardar Vallabhbhai National Institute of Technology, Surat} \\
\textit{$^2$ Faculty of CSE, Sir Padampat Singhania University, Udaipur, India }\\
supriya.mishra@spsu.ac.in}

}

\maketitle

\begin{abstract}
Modern databases typically makes use of the Log Structured Merge-Tree for organizing data in indexes, which is a kind of disk-based data structure. It was proposed to efficiently handle frequent update queries (also called update intensive workloads) databases. In recent years, LSM-Tree has gained popularity and has been adopted by a number of NoSql databases, and key-value stores. Since LSM-Tree was first proposed, researchers and the database community started efforts to improve different components of LSM-Tree. In recent years, Non-volatile Memory, also called Persistent Memory, has also gained significant popularity. This is a class of memory that is non-volatile and byte-addressable at the same time, and hence also termed Storage Class Memory. Apart from that, storage class memory exhibits the combination of the best characteristics of both memory and storage. An overview of the current state of the art in LSM-Tree-based indexes, data systems, and Key-Value stores is provided in this paper.
\end{abstract}

\begin{IEEEkeywords}
LSM-Tree, Database indexing, Disk-based indexing, Data Syatems, KV-stores
\end{IEEEkeywords}

\section{Introduction}
The Log Structured Merge-Tree (LSM-Tree) \cite{lsm} paper was the first one to present the architecture to facilitate low-cost indexing for update-intensive workloads in databases. Unlike traditional indexes that perform an in-place update, LSM-tree performs an out-of-place update. An in-place update, updates the underlying data structure on disk immediately, incurring a random disk I/O. In an out-of-place update, several updates are buffered in main memory and periodically the accumulated updates are flushed to disk, which incurs a sequential disk I/O. In LSM-Tree, the idea is to have an in-memory structure termed as memtable, which serves the insert, delete, and update queries, once the memtable is full, transfer it to the in-disk structure termed as Sorted String Table (SStable).

One of the distinguished characteristics of LSM-Tree is its tunable structure. The levels and the components can be added on demand. The merge operation keeps reorganizing the data for better space utilization and its operation's performance within the theoretical bounds on write amplification. Additionally, LSM-Tree design brings several advantages, such as better write performance, better space utilization, simplified concurrency, and high recoverability. Leveraging these benefits, Facebook had widely adopted LSM-based storage in RocksDB \cite{rocks}, which can handle real-time data processing \cite{realtime}, stream processing \cite{realtime}, graph processing \cite{graphprocessing}, and online transaction processing workloads.

In recent years, many open-source NoSql databases incorporating LSM-tree have been proposed by the research community. The LSM-Tree is the most commonly utilized structure by today's NoSQL systems \cite{lsmsurvey}, including BigTable\cite{bigtable}, Dynamo [24], HBase\cite{hbase}, Cassandra \cite{cassandra}, LevelDB\cite{leveldb}, RocksDB\cite{rocks}, and AsterixDB\cite{storageAsterixdb}. With the availability of novel Non-volatile Memory (NVM), also called Persistent Memory (PM) or Storage Class Memory (SCM), researchers started utilizing PM\footnote{In this paper, we use the term ``PM" to refer to NVM, SCM, and PM altogether.}. PM has some main memory like features i.e low latency, byte-addressability and some persistent storage like features such as large capacity, and non-volatility [8]. This paper presents a comprehensive assessment of the most recent advancements in LSM-Tree based indexes, LSM-Tree based storage systems, and KV-stores. The contributions of this paper are as follows-
\begin{itemize}
    \item We present the details of LSM-Tree, and essential concepts of novel PM.
    \item We present a literature review of state-of-the-art in LSM-Tree based indexes, data systems, and KV-stores.
    \item We summarise the key characteristics of the surveyed papers.
    \item In the end, we present a few research gaps and potential future directions, that we have identified from our survey.
\end{itemize}

The rest of the paper is organized as follows. Section 2 presents the background of LSM-Tree index, its components and working. This section also includes the basics of PM. Section 3 presents a review of LSM-Tree based indexes, data systems, and KV-stores. Section 4 presents the conclusion and the opportunities for future work.

\section{Background}
In this section, we discuss the basics of LSM-Tree. We discuss the novel PM technology in brief.
\subsection{LSM-Tree: Structure and Working}
An LSM-Tree is a hierarchical, sequential, and disk-based indexing structure that supports common index operations such as insert, delete, update, and search. It is capable of storing enormous amounts of data and optimized for handling update queries\footnote{This study uses the phrase "update queries" to include the regularly used insert query, update query, and delete query.} generated at a very high rate, by today's applications like Social Media, Location-based Services, etc.
\newline
\newline
\textbf{LSM-tree Architecture:} LSM-tree design is composed of a combination of in-memory and in-disk data structures. LSM-Tree has a smaller in-memory data structure which is memory resident called as \textbf{Memtable} and a number of in-disk data structures called as \textbf{SStables}. In addition to Memtable and SStables, there are some other components in the architecture namely, Bloom filter, Write ahead log (WAL), and Compactor, as shown in figure \ref{fig:lsm}. All the components of the architecture are described as follows: 
\begin{figure}
	\centering
	\includegraphics[width=\columnwidth]{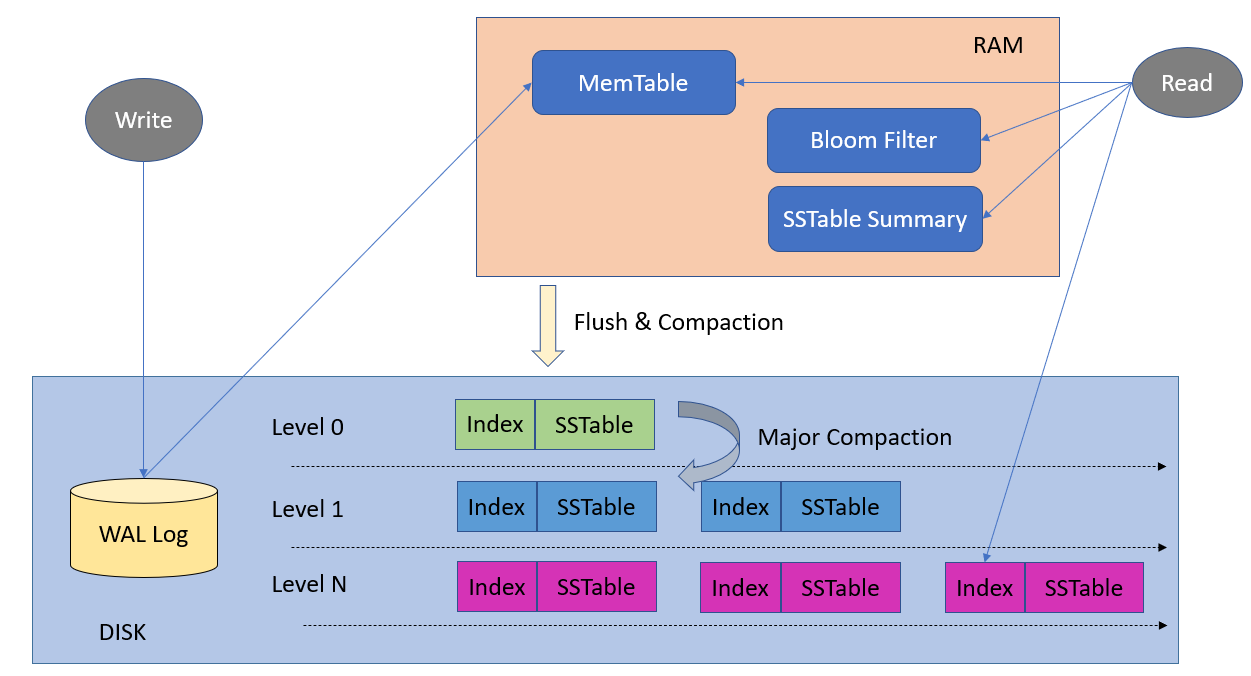}
	\caption{ Architecture of LSM-Tree} 
	\label{fig:lsm}
\end{figure}
\begin{itemize}
 \item \textbf{Memtable:}
        This is an in-memory structure maintain in DRAM. This is used as a buffer for handling upcoming updates (insert/delete/update). Any data structure can be used for memtable implementation but height balanced trees, skiplists are used majorly. All operations are performed on memtable. For insert/update query, before writing the data to memtable, it is written in write ahead log (WAL) file. For delete query, a special marker for key is added to memtable to indicate the key is deleted.
        
        \item \textbf{SStable:} Memtable has a limited size, therefore, once it is full, the memtable's data flushed to disk component in sorted order, which is called an Sorted Strings table (SStable). The flush operation uses sequential disk I/O, which is the reason for high performance. The sorted data in SStable benefits search query. 

        \item \textbf{Bloom filter:}
        Bloom filter is used to check if the given search key is present in the memtable in O(1) time. It is a probabilistic data structure, there are changes of false positives matches, but false negatives matches does not happen. In cases when key does not exist, bloom filter results a false, so there is no need to search that key in memtable, and search query can output "false/Key not found". Hence, bloom filter improves search query performance.   

        \item \textbf{WAL:}
        A write-ahead log called WAL is used to ensure that data is preserved even in the event of a system failure. Before updating the in-memory data structure in response to a write request, the data is first added to a WAL file, also known as a journal, and flushed to the disk using direct IO. In the event that a system crashes before persisting the in-memory data structure to disk, this enables systems to recover from the WAL.

        \item \textbf{Index:}
        Along with each SStable, a small index is maintained, that is used to provide lookup for the particular SStable. 
        
        \item \textbf{Compactor:}
        As a result of flushing Memtable to SStable, their are some obsolete entries in the SStables. The valid copy is the one present in the most recent SStable, but obsolete copies might be scattered in all SStables. The compactor is a background process which runs periodically and merges two or more previous SStable to remove the deleted and redundant entries. This improves search query performance and space utilization.
        
    \end{itemize}

\subsection{Persistent Memory: Characteristics and Challenges}
A number of PM technologies have emerged recently, namely, Phase Change Memory (PCM), Resistive-RAM (Re-RAM), Memrister, Spin Transfer Torque-RAM (STT-RAM), and 3D-Xpoint. Intel has released ``Intel Optane DCPMM" in year 2019, a hardware for PM, since then research has progressed a lot. The key characteristics and challenges of PM are described as follows:
\begin{enumerate}
    \item \textbf{Characteristics:} The key characteristics of PM are as follows-
    \begin{itemize}
    \item \textbf{Byte-addressablity:} Like DRAM, PM is directly accessed by CPU using load/store instructions. It is connected with the CPU with the same bus which is used to connect DRAM. This feature refers to byte-addressability in the context of PM.
    \item \textbf{Durability:} Unlike DRAM, the data stored in PM is durable/non-volatile and remains during power failures.
    \item \textbf{Access Latency:} PM offers read/write latency that is comparable to DRAM. 
    \item \textbf{Cost:} The cost of PM is much lower than PM and slighly more than the storage devices.
    \item \textbf{Large Capacity:} Unlike DRAM, PM offer large memory capacity.
    \end{itemize}
    \item \textbf{Challenges:} The challenges with PM are as follows-
    \begin{itemize}
    \item \textbf{Crash consistency:} The data once stored in PM persists after a crash. If a crash happens in the middle of a store instruction then the crash can cause inconsistency in data. This needs to be handled by the programmer.
    \item \textbf{Data reordering:} When multiple cache-lines are written to PM, then the write may not follow the order of the instructions issued by CPU, which causes a data reordering problem. 
    \item \textbf{PM leaks:} If a crash happens after allocating the PM object and before that object is written to the memory then this situation causes a PM leak problem.
    \end{itemize}
\end{enumerate}

\section{LSM-trees: Review:}
In this section, we present a review of state-of-the-art LSM-Tree based indexes, data systems, and KV-stores.
\subsection{LSM-Tree based Indexes:}

sLSM \cite{slsm}, maintains a number of skiplists to implement memtable. These are called as, “Runs”, denoted by R, and only one of the run is active at any time given. The disk-based store consists of a hierarchical storage system that exhibits a consistent scaling factor for each tier. The data in the disk is maintained in a number of levels, denoted by L, and each level has a number of runs, denoted by D. The data in each of D runs is sorted. When the number of runs at any level reaches D, one more level is added, and so L is incremented by 1. The authors present a merging method based on heaps, which exhibits a time complexity of O(n log(mD)) and a space complexity of O(mD), where n denotes the number of components being merged.

LSM-Tree is not efficient for secondary indexes, first, due to the key is scattered among multiple levels, and second, secondary key can have multiple values, hence maintain obsolete entries is an overhead. The authors of Perseid \cite{perseid} have proposed an efficient secondary index for novel Persistent Memory (PM). PM is a class of memory which is non-volatile and byte-addressable at the same time. The design of Perseid addresses the above mentioned issues. The key features of the design are, (1) PM-DRAM hybrid memory layout, (2) hash-based validation to filter out the obsolete entries, and (3) optimization for insertion and query operations. They have evaluated their results along with FAST\&FAIR, and LSMSI(LSM-based index in leveldb++) on YCSB workloads. Perceid out performs PM-based indexes by 3-7× and achieves about two orders of magnitude performance of state-of-the-art LSM-based secondary indexing techniques even if on PM instead of disks.

Wang et. al \cite{dlsm} proposed an LSM-Tree based index for disaggregated memory environments. The disaggregation memory environment is the one where CPU and memory are physically separated and linked using fast network based interconnection such as, RDMA. They have leveraged the architecture of LSM-Tree for the fact that memtable and SStable of LSM-Tree are inherently disaggregated and hence experimented the feasibility of LSM-tree based index with memory disaggregation. The key features of their work are novel index design for disaggregated memory architecture, reduced synchronization overhead, operations optimized for byte-addressable memory. The data in PM nodes is written sequentially. To ensure consistency of operations, they have proposed a transaction commit algorithm, named “Reorder Ring”(ROR). With experiments, they have evaluated that utilizing PM gives 3.8x gain in YCSB benchmark and 2x gain in TPC-C benchmark.

\cite{nvcache} is non-volatile cache for LSM-Tree based indexes. They have introduced one layer between DRAM, which stores memtable, and SSD/HDD which store SSTables. This new layer is called as NV-cache. There are two specific benefit of this design, first, the gap between write latencies of DRAM and SDD is bridged by using PM-based cache in the middle, second, write amplification of compaction algorithm is reduced. NV-cache architecture, the memtables are stored inside the DRAM, once the memtable is full, it can be quickly flushed to byte-addressable PM-based cache (NV-cache). NV-cache utilizes skiplist to store the flushed data, for efficient compaction. To flush data from NV-cache to SSD, they have proposed LRU(Least Recently Used)-based cache eviction mechanism. With their experiments on a number of KV-stores, like, LevelDB, NoveLSM, and MatrixKV, using db\_becnh and YCSB workloads, they achieved 2.3x write reduction and upto 54\% improvement in performance.

The LSM-Tree based PM system is redesigned by NoveLSM \cite{novelsm}. It points out a few main problems in modifying the current LSM-Tree systems to take advantage of PM. First, a mismatch in the data format between the Sorted-String Table (SST) and memtable can result in a large serialization and de-serialization cost in the LSM-Tree. NoveLSM uses an in-memory mutable memtable and an in-PM immutable memtable to overcome this. The data is flushed into the immutable memtable when the changeable memtable fills up. Second, when the memtable is full and compressed to disk, the LSM-Tree may induce a write stall because it is the only memtable that can be written into or changed in-memory. NoveLSM uses PM's byte-addressability and permits in-place updates in the PM resident memtable to address this. 

A tiering LSM-Tree based persistent index is called TLSM \cite{tlsm}. It uses the three memory hierarchy tiers—DRAM, PM, and HDD, to modify the LSM-Tree design for PM (simulated as NVRAM). PMDK is used to replicate PM on DRAM. Write stall and write amplification are two problems with the current LSM-based systems that the authors have addressed. They addressed the problems and investigated whether PM might be used as a middle-layer store. In order to minimize compaction IO overhead in the disk, a persistent skip list is built using byte-addressable compaction in the middle layer of the DRAM, which stores changeable and immutable memtable data. When data is flushed from PM to HDD, the three tiering policies they suggested are used. After a thorough study, we summarize the key points of the reviewed papers in Table \ref{comparison1}.

\begin{table*}[]
\centering
\caption{Comparison of LSM-Tree based Data Systems and KV-stores}
\label{comparison1}
\resizebox{\textwidth}{!}{%
\begin{tabular}{llll}
\hline
\textbf{Paper}                           & \textbf{Year} & \textbf{Description}                                                                                                                       & \textbf{Remarks}                                                                        \\ \hline
sLSM \cite{slsm}        & 2018          & \begin{tabular}[c]{@{}l@{}}-Used skiplist data structure to implement memtable \\ -Implemented Min-Heap based merge operation\end{tabular} & \begin{tabular}[c]{@{}l@{}}-Merge complexity O(n log(mD))\\ -Can not be used for secondary index\end{tabular} \\
NoveLSM \cite{novelsm}  & 2018          & \begin{tabular}[c]{@{}l@{}}-First LSM-based index to Utilize PM\\ -Uses DRAM-SSD for storing memtable \\ and SStable respectively \end{tabular}  & \begin{tabular}[c]{@{}l@{}} Reduced serialization/deserialization cost\\-Faster compaction \\-Reduced write stalls caused by LSM-Tree \\ Not cost effective for large data system as all the data is stored inside PM\end{tabular}                                                                                   \\ 
TLSM \cite{tlsm}        & 2020          & \begin{tabular}[c]{@{}l@{}}-Enhanced NoveLSM Design\\-Utilized DRAM-PM-SSD memory hierarchy for storing data\end{tabular}                                                                                                                                         & -Improved compaction and merge algorithms as compared to NoveLSM                                                                                        \\
NV-cache \cite{nvcache} & 2022          &   \begin{tabular}[c]{@{}l@{}}-Added PM based cache between DRAM and SSD \\- Used Skiplist to implement cache for better compaction \end{tabular}                                                                                                                                          &  - Achieved better performance than existing works like, NoveLSM                                                                                     \\
dLSM \cite{dlsm}        & 2023          &    \begin{tabular}[c]{@{}l@{}}- Introduced the usage of disaggregated memory in LSM-Tree design\\- Used RDMA to connect CPU with byte-addressable PM \\- Proposed PM based memory allocator and semi-persistent memtable \end{tabular}                                                                                                                                          &    \begin{tabular}[c]{@{}l@{}}- Achieved performance gain for YCSB and TPC-C benchmarks \\-Reduced synchronization overhead in disaggregated environments \end{tabular}                                                                                     \\ 
Perseid \cite{perseid}  & 2023          &   \begin{tabular}[c]{@{}l@{}} -Utilized DRAM-PM memory hierarchy \\Hash-based validation strategy to remove obsolete entries \end{tabular}                                                                                                                                         &     \begin{tabular}[c]{@{}l@{}}-Enhanced query performance \\-Achieved better performance than existing state-of-the-art \end{tabular}                                                                                     \\ \hline
\end{tabular}%
}
\end{table*}

\subsection{LSM-Tree based Data Systems:}
A number of open source NoSql data systems implementing LSM-Tree have been proposed by the research community namely, RocksDB \cite{rocks}, Cassandra \cite{cassandra}, LevelDB \cite{leveldb}, HBase \cite{hbase}, and AsterixDB \cite{asterixdb}. 

\textbf{LevelDB:} It is an open-source data system based on LSM-Tree supporting key-value stores. It was developed by Google in 2011 as an embedded storage engine for higher performance. It supports operations like puts(Insert operation), gets(retrieve operation), and scans(search operation). It implements a leveling merge policy which was adopted by the subsequent works in the field. 

\textbf{RocksDB:}
RocksDB \cite{rocks} is designed by adopting LSM-Tree to optimize space utilization. Initially, it was an extension of LevelDB and later many new features were introduced by Facebook. RocksDB has adopted a leveling merge policy having a size ratio of 10. Its design stores 90\% of the data at the last level and ensures space wastage as low as 10\% of overall space. RocksDB has proposed several improvements to the existing system. To reduce the write amplification, it has proposed a tiering merge policy at level 0 and dynamic resizing of all the levels as per the size of the last level. To optimize the CPU and disk resources consumption by the merge operation, it has adopted a leaky bucket mechanism \cite{leakybucket}.

\textbf{HBase:} 
HBase \cite{hbase} is developed by Apache, and is a distributed data system based on Hadoop. It has a master-slave architecture. It is modeled based on Google’s Bigtable \cite{bigtable}. It dynamically partitions the dataset into multiple regions and LSM-Tree is used to manage the storage within each partition. It implements a tiering merge policy which has two variations:
\begin{enumerate}
    \item Exploring Merge Policy, \item Date-tiered Merge Policy.
\end{enumerate}

In the Exploring Merge policy, a thorough evaluation is conducted on all components that can be merged. The selection is then made based on the component that incurs the lowest writing cost. The merging policy implemented in HBase is referred to as the default merging policy. On the contrary, the Date-tiered merging approach has been specifically designed for the management of time series data. The selection of components is based on their respective time ranges, resulting in the final components being partitioned according to time range.
 In terms of spatial queries, this policy is efficient.

\textbf{Cassandra:}
Cassandra \cite{cassandra} is an open-source distributed data system released by Apache and it is modeled based on Amazon’s DynamoDB \cite{dynamo} and Google’s BigTable \cite{bigtable}. Unlike HBase, it has a decentralized architecture addressing the issue of a single point of failure. Each data partitioned is managed by an LSM-Tree. It implements multiple merging policies like, tiering merge policy, leveling merge policy, and date-tiered merge policy. It provides support for secondary indexes which is a major improvement over existing systems.

\textbf{AsterixDB:}
Apache’s AsterixDB \cite{asterixdb} is an open source Big Data system which aims to manage massive amount of unstructured data. The datasets are hash-partitioned based on their primary keys and distributed across multiple nodes. Each partition consists of a primary index and a number of secondary indexes which are managed by LSM-Tree. It has an LSM-based B+Tree for normal queries, inverted indexes to support full text and similarity queries \cite{similarityquery}, and LSM-based RTree for spatial queries. It provides a framework to convert any in-memory index to LSM-Tree based index, which is termed as LSM-ification.

\subsection{LSM-Tree based KV-stores:}
Key value stores play a primary role in the effective management of data in NoSQL systems. LSM-Tree is the widely used structure in most of the KV stores.

Lu et al. proposed a key-value separation mechanism in WiscKey \cite{wisckey}. It is KV-store that maintains KV pairs in log files stored on SSD and Key is stored in an LSM-Tree in DRAM. This KV separation can decrease the size of the LSM tree, which has several advantages like reduced long-tail latency of the operations. Initially, it lessens the length and frequency of the compaction process, which has an impact on the tail latency. Moreover, this method lessens the LSM-Tree's write amplification, which extends the SSD's life. Thirdly, because little writes may be absorbed by the DRAM and subsequently batched into the SSD, it scales well with high-performance storage devices.

In \cite{gdh}, the authors have proposed a novel method, gradient data hierarchy based on hotness for data for migrating cache data in the storage hierarchy. The major goal of this paper is to improve read/write performance, and improve cache hit rate. To achieve this, they have proposed a novel design and Gradient hierarchy of hot and cold data. Skiplist is a commonly used structure for storing in-memory updates. Instead of using skiplist, they have utilized hashtable and this reduces the time complexity of operation from O(log N) to O(1).

\begin{table*}[]
\centering
\caption{Comparison of LSM-Tree based Data Systems and KV-stores}
\label{comparison2}
\resizebox{\textwidth}{!}{%
\begin{tabular}{llll}
\hline
\textbf{Paper}                           & \textbf{Year} & \textbf{Description}                                                                                                                       & \textbf{Remarks}                                                                        \\ \hline
AsterixDB \cite{asterixdb}        & 2014          & \begin{tabular}[c]{@{}l@{}}-First system to handle big data utilizing LSM-Tree \\ -Implements a number of LSM-Tree based primary and secondary indexes \\ Proposed LSM-ification framework to convert any index to LSM-Tree index\end{tabular} & \begin{tabular}[c]{@{}l@{}}-Open source big data system\end{tabular} \\

RocksDB  & 2017         & \begin{tabular}[c]{@{}l@{}}-Developed by Facebook as an extension of LevelDB\\ -Implemented levelling merge policy \\ Dynamic resizing of levels \end{tabular}  & \begin{tabular}[c]{@{}l@{}}-Open source big data system \\-Highly optimized for space utilization\\ -Implement novel techniques for CPU resouce utilization\end{tabular}                                                                                   \\ 

LSM-Trie \cite{lsmtrie}        & 2015          & \begin{tabular}[c]{@{}l@{}}-LSM-Tree based KV-store\\-Key characteristic huge storage capacity\end{tabular}                                                                                                                                         & -High throughput of write and read operations                       \\

LDS  \cite{lds} & 2017          &   \begin{tabular}[c]{@{}l@{}}-LSM-Tree based direct storage system \\- Proposed novel file system abstraction \\stating the fact that file system level indexes \\are not optimized for LSM-Tree \end{tabular}                                                                                                                      &  \begin{tabular}[c]{@{}l@{}}-With LDS LSM-Tree are able to achieve upto \\ 2.5 times improved performance\end{tabular}                                                                                     \\

WiscKey \cite{wisckey}        & 2017          &    \begin{tabular}[c]{@{}l@{}}- LSM-Tree based KV-store\\- In this, key is stored in LSM-Tree\\, and KV-pair is written to log file stored in SSD\\-Size of memtable is smaller \end{tabular}      &    \begin{tabular}[c]{@{}l@{}}-The key achievement is reduced tail latency \end{tabular}                                                                                     \\ 

SpanDB \cite{spandb}  & 2021          &   \begin{tabular}[c]{@{}l@{}} -KV-store motivated by RocksDB \\-Utilized normal SSDs as well as NVMe SSDs \\for better IO and cost effectiveness \\-Utilized SPDK library for writing into NVMe SSD\end{tabular}                                                                                                                                         &     \begin{tabular}[c]{@{}l@{}}-Achieved improved disk utilization, \\which is the bottleneck in LSM-Tree \end{tabular}                                                                                     \\
NVKVS \cite{nvkvs}  & 2022          &   \begin{tabular}[c]{@{}l@{}} -Addressed multi-threading issues in LSM-Tree based KV-stores  \\Utilised PM to addressed issues of KV separated stores \end{tabular}                                                                                                                                         &     \begin{tabular}[c]{@{}l@{}}-Able to achieve reduced write latency \\-Achieve increased system durability \end{tabular}                                                                                     \\

GDH \cite{gdh}  & 2023          &   \begin{tabular}[c]{@{}l@{}} -Proposed GDH based data migration \\Utilized hash table to implement memtable \end{tabular}                                                                                                                                         &     \begin{tabular}[c]{@{}l@{}}-The time complexity of operations\\ is reduced from O(log N) to O(1)\end{tabular}                                                                                     \\
\hline
\end{tabular}%
}
\end{table*}

In SpanDB\cite{spandb}, the authors have proposed an LSM-Tree based KV store, which adapts RocksDB. They have identified that the disk IO becomes the performance bottleneck for LSM-Tree based systems. They have adopted a hybrid storage organization for storing the bulk of data in cost-effective, high-volume SSDs, called the capacity disk (CD) and Log file, and top levels of LSM-Tree in fast NVMe SSDs, called as speed disk (SD). Like RocksDB\cite{rocks}, all the memtables are maintained inside the DRAM. SpanDB utilizes SPDK (a set of libraries developed by Intel for NVMe-based SSD) for parallel writing of WAL, allows for asynchronous request processing in order to reduce the overhead of inter-thread synchronization and effectively handle polling-based I/O.

LSM-trie \cite{lsmtrie} is a high-capacity Key-Value storage system optimized for efficiently handling multi-terabyte data on a single server. Its innovative features, such as a compaction architecture combining linear and exponential growth, lead to enhanced write throughput and reduced write amplification. LSM-Tree's strength lies in its ability to manage billions of small data items with minimal RAM usage, supporting over 500K writes and 50K reads per second across diverse workloads. This positions LSM-Tree as a potent solution for data-intensive applications that demand high write throughput.

In \cite{lds}, the authors propose a novel LSM-Tree based Direct Storage system to overcome limitations in file system level indexes, leveraging the copy-on-write feature for efficient storage management and simplified consistency control. Through comprehensive experiments comparing LSM-Tree performance with LDS against other file systems on both HDDs and SSDs, the authors demonstrate significant improvements, achieving LSM-Tree write throughput up to 3 times higher on HDDs and 2.5 times higher on SSDs, highlighting the enhanced performance potential of their approach.

NVKVS\cite{nvkvs} is a non-volatile memory Key-Value Store (KV-store) that employs a key-value separation strategy. It addresses the bottleneck associated with KV-separated stores by saving data directly to Non-Volatile Memory (NVM), enhancing write throughputs and reducing data write times. By segregating writer and client threads and utilizing RocksDB for indexing, NVKVS optimizes read operations for efficient and speedy access. In write-intensive scenarios, the system significantly reduces data write times, contributing to improved system resilience.

Now, we summarize the reviewed papers on LSM-Tree based data systems and KV-stores, in the table \ref{comparison2}.

\section{Conclusion and Future Directions}
LSM-Tree is the widely used structure for handling a large number frequent update queries, and hence it is used by many data systems and KV-stores. In this paper, we have presented the state-of-the-art in LSM-Tree based Indexes, Data Systems, and KV-stores. The emergence of novel PM has opened new directions of research and optimizations in currently used LSM-Trees. Compared to DRAM, PM has a much larger capacity and lower cost and power consumption, and hence PM has potential to be mainstream memory in years to come. From our survey, we have identified a few possible future directions:

\begin{itemize}
    \item From our survey, we have identified that majority of LSM-Tree based indexes are not suitable for implementation of secondary index. There is a scope to develop LSM-Tree based secondary indexes that are efficient. 
    \item The involvement of Machine Learning algorithm to develop LSM-Tree based learned indexes, might further enhance the query performance.
    \item eADR enabled PM system, might enhance the performance of existing PM-based LSM-Tree architectures.
    \item Apart from PM connected to DDR buses (like Intel Optane DCPMM), a new type of SCM known as CXL device-attached memory (like Sumsung's Memory-semantic SSD [8]) is introduced by the high-bandwidth and low-latency IO connectivity, Computer Express Link (CXL) [3, 29].
\end{itemize}

\end{document}